\documentclass[floatfix,twocolumn,aps,prb,superscriptaddress]{revtex4-2}
\usepackage[T1]{fontenc}
\usepackage{float}
\usepackage{amsmath}
\usepackage{graphicx}
\usepackage{xcolor}
\usepackage{pdfpages}
\usepackage{pgffor}
\makeatletter
\AtBeginDocument{\let\LS@rot\@undefined}
\makeatother
\usepackage{hyperref}

\begin{abstract}
Spin-active color centers are the basis of solid-state defect systems utilized in quantum technologies. Although silicon is an emerging host material for quantum defects, there is an urgent need to characterize color centers with a non-zero electron-spin ground state in this platform, in addition to the prominent T center. In this work, we perform first-principles calculations to identify the possible atomic structures compatible with the experimentally observed N-line series in silicon. We propose that the core structure of the N1 center consists of a neighboring carbon and nitrogen interstitial pair. Furthermore, we predict that more complex defects involving self-interstitial and interstitial oxygen atoms are feasible candidates for the further lines in the series. As all of these color centers are isoelectronic to the T center, they provide a family of alternative spin qubits with emission near the low-energy telecommunication bands.
\end{abstract}

\begin{document}
\title{First-principles insights into the atomic structure of carbon-nitrogen-oxygen complex color centers in silicon}
\author{P\'eter Udvarhelyi}
\email{udvarhelyi.peter@nims.go.jp}
\affiliation{International Center for Young Scientists, National Institute for Materials Science, 1-1 Namiki, Tsukuba, Ibaraki, 305-0044, Japan}
\maketitle

\section{Introduction}
Silicon is an emerging material platform for defect-based quantum technologies~\cite{Awschalom_2018}. As one of the most technologically mature materials, it offers advantages in growth, microfabrication, and CMOS-compatible device integration~\cite{Bergeron_2020, Wang_2020}. Recent advances in isotope engineering~\cite{Holmes_2021} and ion implantation techniques~\cite{Andrini_2024, Hollenbach_2025} have positioned silicon as a promising host for quantum defects, joining the more established platforms of diamond and silicon carbide~\cite{Katsumi_2025, Zhou_2025}.

The family of carbon-related quantum defects is the focus of recent experimental and theoretical investigations in silicon. Several such emitter defect centers have been identified that operate in the telecommunication wavelength ranges. Some notable defects are the interstitial carbon defect ($\text{C}_\text{i}$)~\cite{Thonke_1987, Deak_2024}, the G center ($\text{C}_\text{s}\text{C}_\text{i}$)~\cite{Song_1990, Udvarhelyi_2021}, the C center ($\text{C}_\text{i}\text{O}_\text{i}$)~\cite{Davies_1989, Udvarhelyi_2021}, and the T center ($\text{C}_\text{s}\text{C}_\text{i}\text{H}_\text{i}$)~\cite{Safonov_1996, Dhaliah_2022}, where s and i subscripts label substitutional and interstitial defect positions, respectively. The latter is the prominent defect possessing a doublet electron spin ground state~\cite{Bergeron_2020, Higginbottom_2022, Islam_2024}. Despite its favorable spin and optical properties, the involvement of a single hydrogen atom, which passivates one of the two carbon dangling bonds in the color center, makes its deterministic creation challenging~\cite{Dhaliah_2022}. Recent theoretical studies proposed alternatives to circumvent this problem by searching for isoelectronic defect structures to the T center. In these approaches, the CH part of the CCH defect is replaced with a trivalent atom from the boron group (B, Al, Ga, and In)~\cite{Xiong_2024}. Building on previous calculation results on carbon-nitrogen defect complexes~\cite{Platonenko_2019, Kuganathan_2023, Sgourou_2024}, it was recently proposed that nitrogen can work as a simple replacement as well~\cite{Nangoi_2026}. The latter work predicted the $\text{C}_\text{s}\text{N}_\text{i}$ defect in its split-dumbbell configuration as a novel, T-center-like telecom emitter in silicon, despite its larger formation energy compared to its dissociated neighbor form. 

As previous calculations could not establish an unambiguous connection between the proposed defect structures and experimental signals, we aim to systematically map the carbon-nitrogen impurity interaction in silicon via calculations constrained by experimental properties. Here, we focus on identifying the atomic origin of the N-line series in silicon, labelled by the N1, N2, N3, N4, and N5 zero-phonon lines (ZPL), which are positioned closely at 745.6, 758.0, 761.5, 767.4, and 772.4 meV, respectively~\cite{Dornen_1987, Dornen_1988}. The lines were observed only after carbon and nitrogen co-implantation, suggesting the incorporation of these two elements in the centers. Comparing the FZ and CZ silicon samples, oxygen incorporation is restricted only to the N3, N4, and N5 lines. Isotope shift measurements of the ZPL lines are consistent with the involvement of a single nitrogen and a single carbon atom in the core of the defect centers. Thermal annealing studies of the photoluminescence identified the N1 center as the most stable in the set, followed by N2, while oxygen-related lines are the first to vanish at elevated temperature. Experiments focusing on the N1 center revealed that ion implantation is crucial for the formation of the defect, as the line cannot be observed in nitrogen-doped and electron-irradiated samples~\cite{Dornen_1986}. They also reveal the bound-exciton nature of the optical transition and the Zeeman-splitting of the N1 line, consistent with a spin-doublet electron state. Consequently, the color center is in its neutral charge state, and the spin originates from the odd number of electrons in the C-N complex. The N1 transition couples to two local vibrational modes (LVM) at 71.3 meV and 122.9 meV with respect to the N1 ZPL energy~\cite{Dornen_1985, Dornen_1986vib}. Furthermore, an exceptionally strong optical coupling of the defect center to the $\text{O}^\Gamma$ vibronic band is observed. Uniaxial stress measurements on the ZPL shift revealed the point symmetries of $\text{C}_{1\text{h}}$ for N1 and N5, and $\text{C}_{1}$ for N2 and N3, respectively~\cite{Dornen_1989}. The stress responses also show a slight change between the N1 and N2 core structures and a smaller perturbation from oxygen. It is concluded that the color centers consist of two slightly different C-N cores, N1 and N2, and that their perturbation by additional oxygen incorporation creates N5 and N3, respectively. There are no further experimental findings available for the N4 line of the series owing to its relatively weak photoluminescence intensity. Although there is strong experimental evidence on the composition and point symmetries of the N centers, their exact atomic configuration and electronic structure remain unidentified.

In this work, we demonstrate that the atomic structures of C-N, C-N-Si, C-N-O, and C-N-Si-O defect stoichiometries showing the lowest formation energy and largest binding energy possess electronic and optical properties that make them possible candidates as the origin of various color centers in the N-series. Moreover, these color centers are isoelectronic to the T center, and therefore share its electron-spin doublet ground state.

\section{Methods}
We employ density functional theory (DFT) calculations with a plane-wave basis and periodic supercells as implemented in the Quantum Espresso code~\cite{QE1, QE2, QE3}. The core electrons are treated in the pseudopotential method using the Optimized Norm-Conserving Vanderbilt (SG15 ONCV) potentials~\cite{Hamman_2013, Schlipf_2015}. GGA density functionals of Perdew-Burke-Ernzerhof (PBE)~\cite{PBE} and the hybrid functional of Heyd–Scuseria–Ernzerhof (HSE06)~\cite{HSE} were used for configuration pre-screening and final results, respectively. The optimized lattice constant of the cubic Bravais cell of the silicon host is 5.477~\AA~and 5.446~\AA~using PBE and HSE06 functionals, respectively.
The calculations are performed in a 216-atom, $3\times3\times3$ silicon supercell using $\Gamma$-point sampling and an 80 Ry plane-wave cutoff.
Convergence tests for the zero-phonon line (ZPL) transition energies were performed in a 512-atom, $4\times4\times4$ silicon supercell using the HSE06 functional with a reduced plane-wave cutoff of 60 Ry to mitigate the computational overhead of this large cell. 
Ionic relaxation with a force threshold of 0.001 a.u. was carried out in all DFT calculations.
Excited states were calculated and relaxed using the $\Delta$SCF method of fixed occupations~\cite{Gali_2009}. Total energies in charged supercells were corrected using the method of Freysoldt {\it et al.}~\cite{Freysoldt_2009}.
Formation energies at charge state $q$ were calculated as
\begin{equation}
E_{\text{form}}^{q}=E_{\text{defect}}^{q}-E_{\text{host}}-\sum_{i}\mu_i n_i +qE_{\text{Fermi}}\text{,}
\end{equation}
where $\mu_i$ is the chemical potential derived from silicon, diamond, $\text{N}_2$, and $\text{O}_2$, while $n_i$ is the number of defects of species $i$ created in the host material.
Optical transition dipoles ($\mathbf{\mu}$) were calculated in the ground state configurations using time-dependent density functional theory (TDDFT) within the Tamm-Dancoff approximation and using the PBE functional~\cite{PBE}, as implemented in the WEST code~\cite{Jin_2023}. Optical lifetimes ($\tau$) from the transition dipole moments are calculated using the Wigner-Weisskopf formula~\cite{Weisskopf_1997}
\begin{equation}
    \frac{1}{\tau}=\frac{2(2\pi)^{3}n_r E_{\text{ZPL}}^{3}\left|\mathbf{\mu}\right|^2}{3\varepsilon_0 h^4 c^3},
\end{equation}
where $n_r=3.42$ is the refractive index of the silicon crystal, $E_{\text{ZPL}}$ is the experimental ZPL energy of the associated optical transition, $\varepsilon_0$ is the vacuum permittivity, $h$ is the Planck constant, and $c$ is the speed of light. A volume-dependent scaling factor of $V/\pi a_{0}^{3}$ was applied to the calculated $\left|\mu\right|^{2}$, where $a_{0}=12.6~\text{\AA}$ is the Bohr radius of the heavy hole in the valence band~\cite{Nangoi_2026}.
Phonon properties of the defects were calculated in the 216-atom supercell using density functional perturbation theory (DFPT) at the PBE level in Quantum Espresso. Localized vibration modes were identified by calculating the inverse participation ratio (IPR) of atomic vibration normal modes $\mathbf{u}_i$ using
\begin{equation}
\text{IPR}=\frac{1}{\sum_{i} \left(\left|u\right|^2_i\right)^{2}} \text{.}
\end{equation}
The partial Huang-Rhys factors ($S_i$) are calculated using the ionic position diffrences from the HSE ground ($\mathbf{R}_{g,I}$) and excited states ($\mathbf{R}_{e,I}$), projected on the phonon modes (($\mathbf{e}_{i,I}$))
\begin{equation}
Q_i=\sum_{I}\sqrt{m_I}\mathbf{e}_{i,I}\cdot\left(\mathbf{R}_{e,I}-\mathbf{R}_{g,I}\right) \text{,}
\end{equation}
\begin{equation}
S_i=\frac{\omega_i}{2\hbar}Q_i^2 \text{,}
\end{equation}
where $\omega_i$ is the normal frequency of the corresponding normal mode $i$. The Debye-Waller factor (DWF) is calculated as
\begin{equation}
\text{DWF}=\text{exp}\left(-\sum_i S_i\right)\text{.}
\end{equation}
The phonon sideband is simulated from $S_i$ using the generator function approach of Alkauskas {\it et al.}~\cite{Alkauskas_2014}.
Atomic structures are visualized using the VESTA code.

\section{Results}
To unveil the atomic structure of the N-lines in silicon, we systematically investigate the interactions between atomic species associated with experimental observations. First, we focus on formation and thermal stability. We search for defect configurations with the most favorable energetics as candidates for further investigations. By gradually increasing the complexity of the models, we explore interactions among C-N, C-N-Si, C-N-O, and C-N-Si-O. Finally, we compare the detailed calculations on the electronic and optical properties of the most stable defect configurations to previous experiments.

\subsection{Aggregation of neutral carbon and nitrogen atom pairs}

Carbon-related impurities are the principal constituents of the recently studied color centers in silicon. Although their aggregation has been studied in detail~\cite{Deak_2023}, their interaction with nitrogen defects requires deeper analysis. Given that the base structure of the N centers consists of a single carbon and a single nitrogen atom, we focus on these defect pairs occupying various sites in the silicon lattice. As the bound exciton transitions and the doublet spin observed in experiments are compatible with a neutral ground state, and both carbon and nitrogen have a large window of stability in their neutral charge states in silicon (see Fig. S1 in Sec. I of the Supplemental Material~\cite{SM} and references~\onlinecite{Bean_1970, Watkins_1976, Davies_1987} therein), we constrain our investigations to neutral products and neutral constituents in binding energy calculations, formulated as
\begin{equation}
E_{\text{binding}}=E_{\text{product}}^{0}-\sum_i^{n} E_{\text{constituent}, i}^{0}+(n-1)E_{\text{host}}\text{,}
\end{equation} 
where negative binding energy corresponds to a stable product complex.
This assumption is confirmed by the convergence of the calculated equilibrium states along the sampled configurations and the small variation of the binding energy when charged constituents are considered (see Fig. S2 in Sec. II of the Supplemental Material~\cite{SM}). The only exception is the C-N-Si interaction, where several structural motifs fall in a close range of formation energy and stability (see Sec.~\ref{sec:motif} for more detail). 

\begin{figure}[t]
    \centering
    \includegraphics[width=0.95\linewidth]{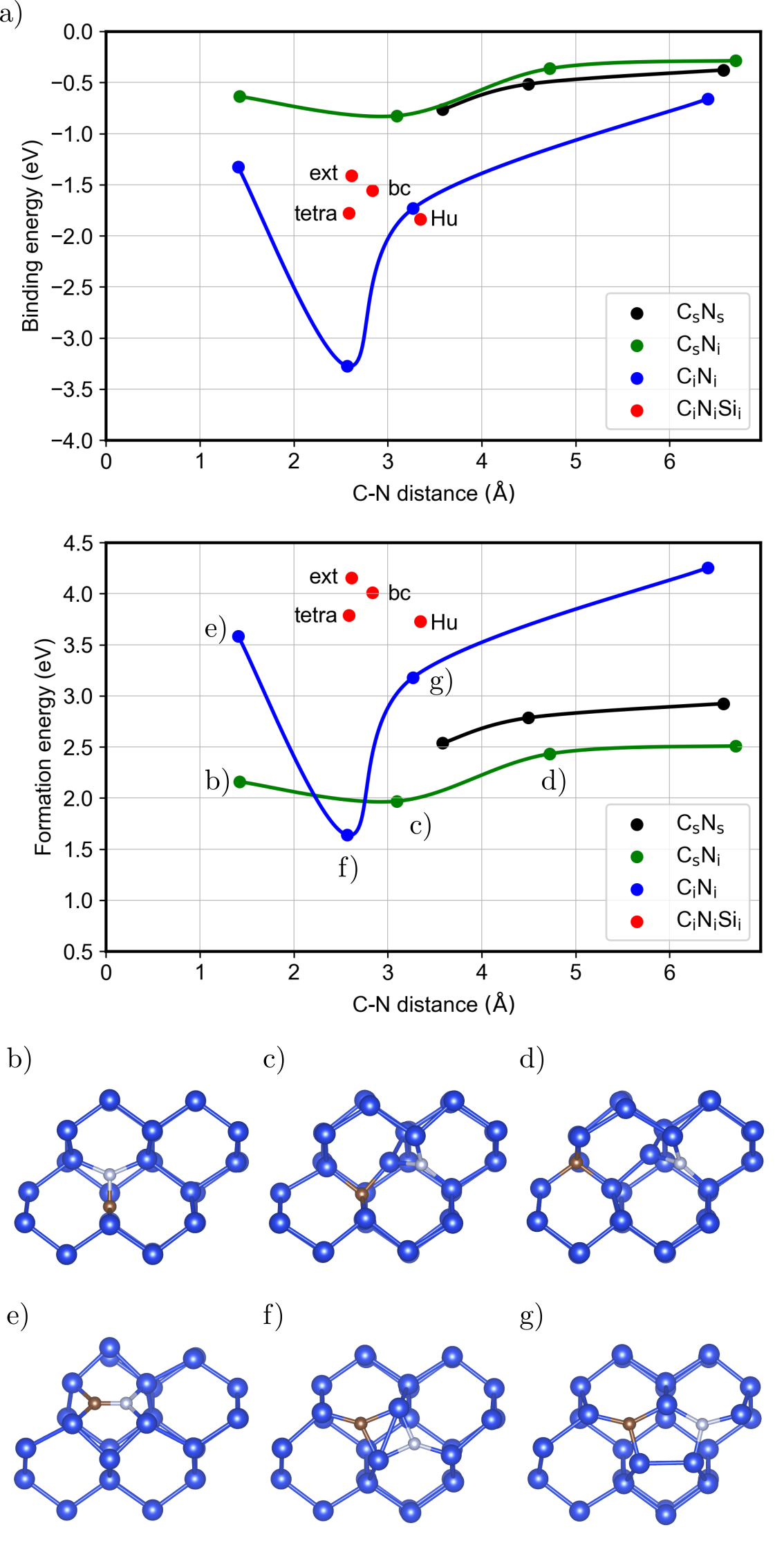}
    \caption{ a) Binding and formation energy plots of carbon and nitrogen interactions in silicon as a function of the interatomic distance. The lines connecting the calculated points at their equilibrium geometries are linear interpolations that guide the eye. b)-g) Corresponding atomic configurations labelled in subfigure a), where brown, dark blue, and light blue balls represent carbon, silicon, and nitrogen atoms, respectively.
For the three-center interstitial defects of C-N-Si (red), only the lowest energy datapoints are included for each defect motif (see Sec. \ref{sec:motif}). The corresponding structures of the latter are visualized in Fig.~\ref{fig:motif}.}
    \label{fig:aggregation}
\end{figure}

Bond reconstruction is expected to yield the most favorable binding energies for interstitial defect pairs, so our sampling strategy focuses on compact configurations that aggregate within the same defect plane and along a single configurational coordinate. This constraint is physically motivated by the bond‑switching diffusion mechanism of split‑dumbbell interstitials, which migrate along the [110] zig‑zag directions. Because the silicon dangling bond defines the diffusion plane, out‑of‑plane approaches hinder reconstruction and do not yield low‑energy bound states. For this reason, restricting the carbon–nitrogen aggregation search to configurations sharing a common [110] axis provides the most realistic route to identifying the energetically preferred defect geometries. Owing to the higher symmetry and fewer dangling bonds of a substitutional defect, it is meaningful to compare carbon-nitrogen pair aggregation along the same [110] direction without losing generality.
The resulting binding and formation energies are plotted in Fig.~\ref{fig:aggregation} a) as a function of the carbon-nitrogen distance in the defect complex. The calculation points correspond to equilibrium defect configurations, while the connecting interpolated curves are shown to guide the eye towards the local minimum points. Focusing only on the carbon-nitrogen structures first, a larger formation energy of the $\text{C}_\text{s}\text{N}_\text{s}$ complex is observed compared to the $\text{C}_\text{s}\text{N}_\text{i}$ complex. This follows from the formation energy difference between the isolated $\text{N}_\text{s}$ and $\text{N}_\text{i}$ defects. The energetically most favorable configuration of the $\text{C}_\text{s}\text{N}_\text{s}$ defect is the completely aggregated, neighboring pair in $\text{C}_{3\text{v}}$ symmetry. The most compact structure of the $\text{C}_\text{s}\text{N}_\text{i}$ complex is the split-dumbbell configuration shown in Fig~\ref{fig:aggregation} b), which was recently predicted as a novel emitter defect~\cite{Nangoi_2026}. However, this configuration is metastable, as its first dissociation step results in the energetically most favorable complex of a neighboring $\text{C}_\text{s}$ and split-dumbbell $\text{N}_\text{i}$ defect, illustrated in Fig.~\ref{fig:aggregation} c). This defect shows no symmetry ($\text{C}_{1}$) consistent with breaking the $\text{C}_{1\text{h}}$ symmetry of the isolated split-dumbbell $\text{N}_\text{i}$ by the adjacent $\text{C}_\text{s}$ defect. Its counterpart, the neighboring $\text{C}_\text{i}\text{N}_\text{s}$ defect, is demonstrated to be unstable, relaxing to the former complex in our calculations.

Finally, we find that the $\text{C}_\text{i}\text{N}_\text{i}$ defect complex is the global minimum configuration in the carbon-nitrogen pair-interaction landscape. Although its dissociated configurations exhibit significantly higher formation energies, this is consistent with the N-series appearing only after ion implantation, where lattice damage can generate high-energy metastable configurations. Despite the hindered formation mechanism, the substantial binding energy of the complex provides a strong drive for aggregation, suggesting a high formation yield after annealing. Analysing the structures along the steps of the dissociation process, illustrated in Fig.~\ref{fig:aggregation} e)-g), we find the most compact metastable bond-centered pair, then at the global minimum, the neighboring split-dumbbell configuration. This latter structure is analogous to the C center, which is a similar aggregated $\text{C}_\text{i}\text{O}_\text{i}$ interstitial defect pair~\cite{Udvarhelyi_2022}. The last notable  $\text{C}_\text{i}\text{N}_\text{i}$ structure is at the next dissociation step, resulting in the second neighbor split pair configuration, where the silicon dangling bonds of the pair reconstruct to form a Humble structure. The visualization and further details of the calculation models are available in Fig. S3 and in Sec. II of the Supplemental Material~\cite{SM}. Based on its exceptional stability among the C-N defect configurations, we select the $\text{C}_\text{i}\text{N}_\text{i}$ complex as the candidate for the N1 core structure, which shows the largest thermal stability among the N-lines in experiments. Next, we search for the structural origin of the N2 line.

\subsection{Motif-based search for complexes involving self-interstitials\label{sec:motif}}

\begin{figure*}[t]
    \centering
    \includegraphics[width=0.95\linewidth]{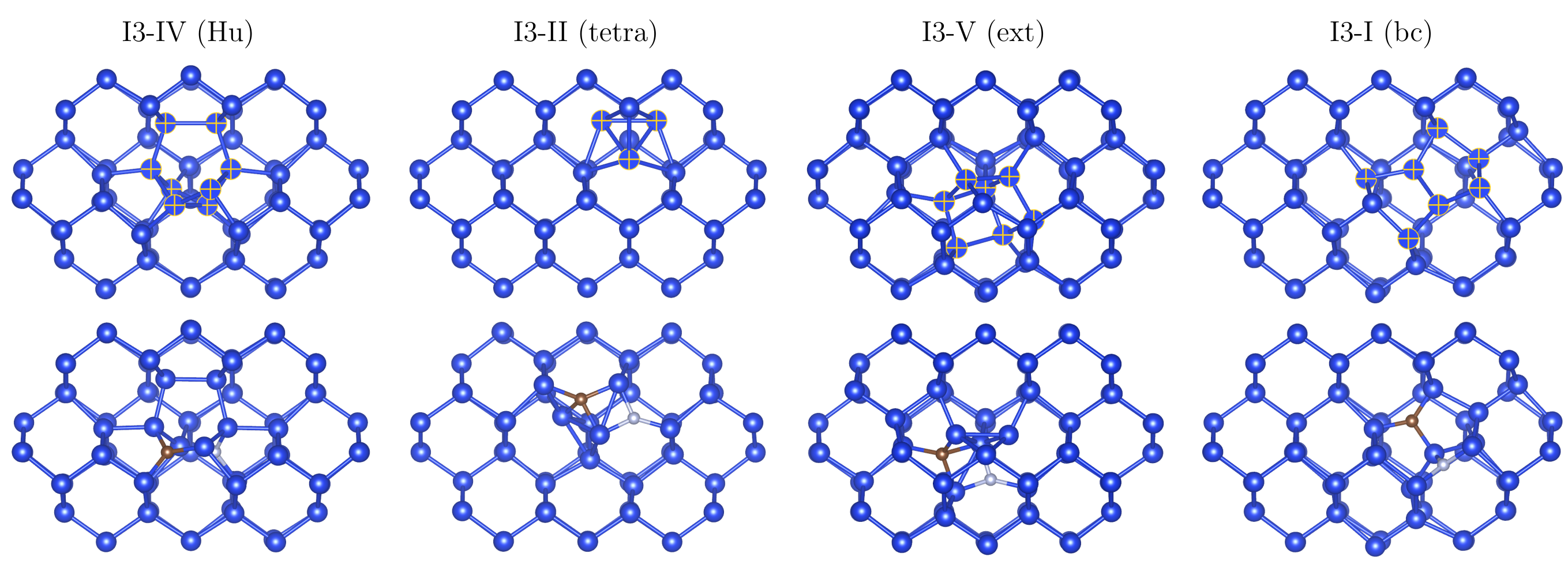}
    \caption{Structural motifs of the lowest-energy tri-interstitial silicon aggregates (upper row) serving as templates for the $\text{C}_\text{i}\text{N}_\text{i}\text{Si}_\text{i}$ complex defect search. The resulting lowest energy configurations are shown under each parent motif. Si, N, and C atoms are drawn in dark blue, light blue, and brown, respectively.}
    \label{fig:motif}
\end{figure*}

As experimental findings suggest that no additional impurity species are involved in the formation of the N2 center, we constrain our search to the involvement of an additional self-interstitial atom. Because of the close energy range of the stable isolated $\text{Si}_\text{i}$ defect configurations (see Fig. S1 in the Supplemental Material~\cite{SM}), the configurational sampling required for the $\text{C}_\text{i}\text{N}_\text{i}\text{Si}_\text{i}$ complex is more complicated than the one-dimensional configurational sampling apllied to the C-N complex. The large variety of the self-interstitial positions is also reflected in their aggregation. Previous calculations on tri-interstitial aggregations identified four low-energy structures~\cite{Carvalho_2005, Santos_2016, Baron_2022}. These are the I3-I bond-centered (bc), I3-II tetrahedral (tetra), I3-IV Humble (Hu), and I3-V extended (ext) shown in the upper row of Fig.~\ref{fig:motif}. Given the complexity of the possible interaction space, we assume that a motif-based search starting from the four lowest-energy self-interstitial aggregate structures, and replacing two silicon atoms in their core with a carbon and a nitrogen atom, will result in the lowest energy structures of the $\text{C}_\text{i}\text{N}_\text{i}\text{Si}_\text{i}$ complex. 
Although this assumption is physically motivated, since the selected motifs provide a wide variety of bonding configurations, we acknowledge that this search strategy is not exhaustive. Consequently, additional low-energy configurations of the $\text{C}_\text{i}\text{N}_\text{i}\text{Si}_\text{i}$ defect complex may exist.
Further details on the initial search for the most favorable configurations and the construction of the model structures in this screening are available in Sec. III of the Supplemental Material~\cite{SM}, where the calculated configurations are visualized in Fig. S4 and the corresponding energies are listed in Table SI. This method indeed yields four compact structures close in energy, as shown in Fig~\ref{fig:aggregation} a). The corresponding structures with the largest binding energies are visualized under their parent I3 defect motifs in Fig.~\ref{fig:motif}. We define the binding energy of the additional self-interstitial defect with respect to the isolated $\text{C}_\text{i}\text{N}_\text{i}$ core structure and the isolated self-interstitial in their neutral charge states. As the "ext" and "tetra" labelled structures result in a large modification of the electronic structure of the $\text{C}_\text{i}\text{N}_\text{i}$ core (see explanation in section~\ref{sec:optical}), we select the bond-centered "bc" and Humble "Hu" structures as promising candidates for the $\text{C}_\text{i}\text{N}_\text{i}\text{Si}_\text{i}$ defect core when interactions with interstitial oxygen is concerned in the next section. 

\subsection{Oxygen interaction with the carbon-nitrogen complex defects}

\begin{figure}[t]
    \centering
    \includegraphics[width=0.95\linewidth]{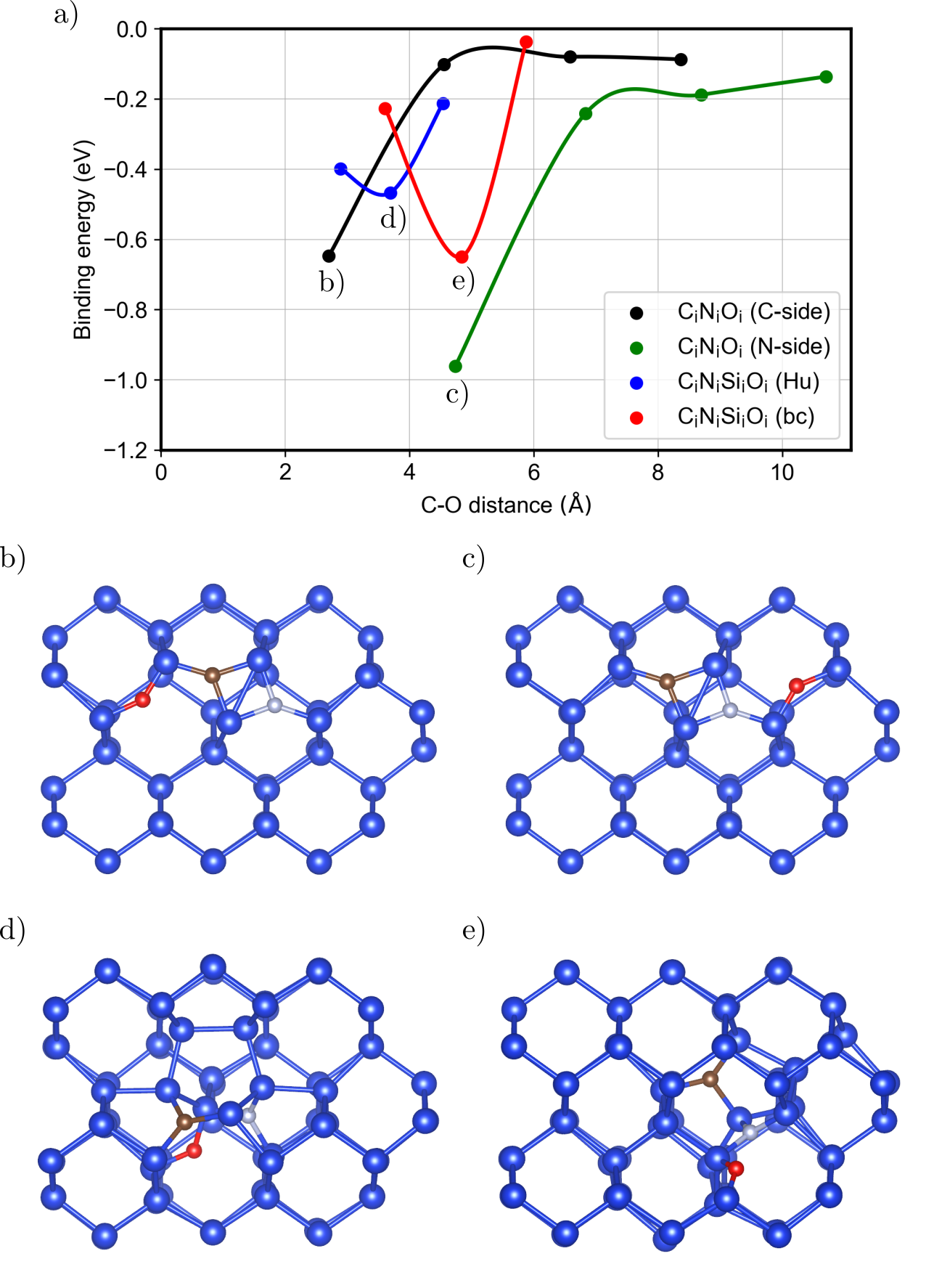}
    \caption{a) Binding energy plot of oxygen interactions with the core structures of selected N1 and N2 candidate defects in silicon as a function of the C-O distance. The lines connecting the calculation points in equilibrium geometries are interpolations to guide the eye along the sampled configurations of the aggregation process. b)-e) Structures with the largest oxygen binding energies in our search. The corresponding data points are labelled in subfigure a). Dark blue, light blue, brown, and red balls represent Si, N, C, and O atoms, respectively.}
    \label{fig:oxygen}
\end{figure}

Strain experiments on the N-lines suggest that the role of oxygen in the N3, N4, and N5 centers is rather a perturbation effect on the corresponding oxygen-free N1 and N2 core structures. Furthermore, oxygen occupies a bond-centered interstitial position in the lattice, making its aggregation straightforward to sample in calculations. We model this by placing $\text{O}_\text{i}$ at all the bond center positions around the core structures of the selected N1 and N2 candidates up to the second neighbor shell. $\text{C}_\text{i}\text{N}_\text{i}\text{O}_\text{i}$ models are generated along the [110] crystal axis and in the (001) plane perpendicular to the C-N core. The lowest-energy oxygen positions preserve the symmetry plane of the C-N core and correspond to nearest-neighbor sites on either side of the core, labelled as C-side and N-side. $\text{C}_\text{i}\text{N}_\text{i}\text{Si}_\text{i}\text{O}_\text{i}$ models are generated by placing the oxygen interstitial atom at all bond-centered positions around the carbon atom up to second neighbor bonds within each $\text{C}_\text{i}\text{N}_\text{i}\text{Si}_\text{i}$ motifs considered here. Selected low-energy configurations are labelled according to their parent motif. Further details are included in Sec. IV of the Supplemental Material~\cite{SM}, where the calculated $\text{C}_\text{i}\text{N}_\text{i}\text{O}_\text{i}$ and $\text{C}_\text{i}\text{N}_\text{i}\text{Si}_\text{i}\text{O}_\text{i}$ configurations and their corresponding energies are shown in Fig. S5 and S6, and Table SII and SIII, respectively. From the calculation results, we identify the lowest-energy configurations and the sampled lowest-energy configurations around them. These selected configurational coordinates are visualized in Fig~\ref{fig:oxygen}, where the binding energy of the additional oxygen is plotted as a function of the C-O distance. We select four structures with large oxygen binding energy, and label them as the $\text{C}_\text{i}\text{N}_\text{i}\text{O}_\text{i}$ (N-side), $\text{C}_\text{i}\text{N}_\text{i}\text{O}_\text{i}$ (C-side), $\text{C}_\text{i}\text{N}_\text{i}\text{Si}_\text{i}\text{O}_\text{i}$ (bc), and $\text{C}_\text{i}\text{N}_\text{i}\text{Si}_\text{i}\text{O}_\text{i}$ (Hu), where "C-side" and "N-side" labels refer to the preferred aggregation direction of $\text{O}_\text{i}$, while "bc" and "Hu" refer to the N2 core structure configurations.

\subsection{Electronic and optical properties\label{sec:optical}}
\begin{figure}[t]
    \centering
    \includegraphics[width=0.95\linewidth]{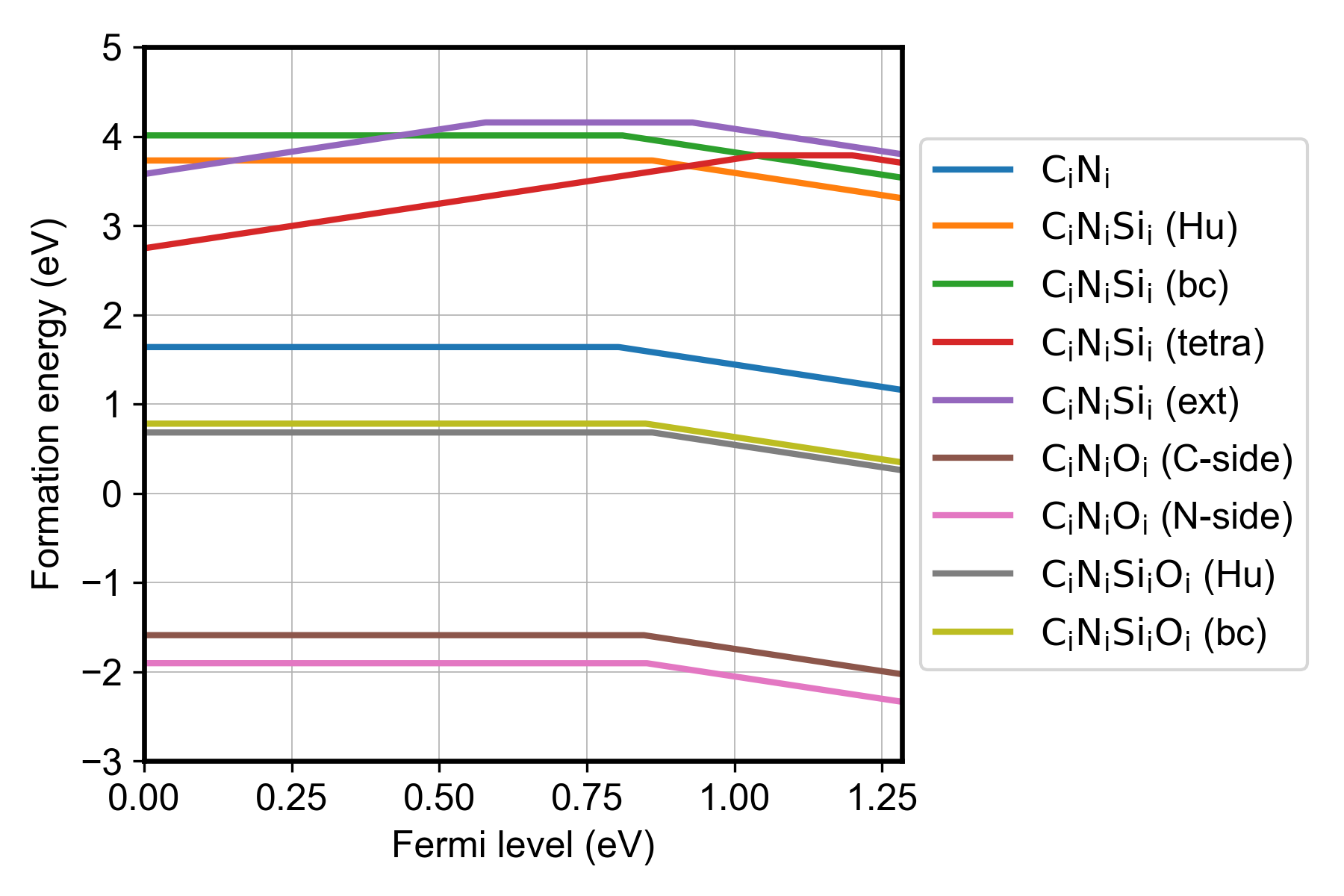}
    \caption{Calculated formation energy diagram in the silicon bandgap for the candidate defects with the largest binding energies. The search for candidates with a single stable transition level compatible with pseudo-acceptor bound exciton centers rules out $\text{C}_\text{i}\text{N}_\text{i}\text{Si}_\text{i}$ (tetra) and (ext) configurations.}
    \label{fig:complex_form}
\end{figure}

After the most stable defect candidates are identified, as evidenced by their large binding energies, their filtering is continued based on their electronic-structure compatibility with the N-lines. The calculated formation energy diagram in Fig.~\ref{fig:complex_form} shows several structures with only a single stable acceptor level compatible with the bound-exciton origin of the N centers. However, $\text{C}_\text{i}\text{N}_\text{i}\text{Si}_\text{i}$ (tetra) and (ext) structures with both donor and acceptor levels active are ruled out, as this would lead to competing exciton transition processes in the color centers. {\it A priori}, we assign the $\text{C}_\text{i}\text{N}_\text{i}$ defect, the energetically most favorable, simplest core structure of the N-series, as the origin of the N1 line. From the remaining two contenders for the N2 center, we select $\text{C}_\text{i}\text{N}_\text{i}\text{Si}_\text{i}$ (Hu) as its charge transition level position is at higher energies compared to the $\text{C}_\text{i}\text{N}_\text{i}$ defect, in-line with the higher transition energy of the N2 line. The oxygen-containing candidates considered here all possess electronic structures compatible with the bound exciton transition.

Next, we focus on the optical properties of the $\text{C}_\text{i}\text{N}_\text{i}$ defect. Its calculated ZPL energy of 701 meV slightly underestimates the experimental N1 ZPL at 745.6 meV. The calculation in a larger 512-atom ($4\times4\times4$) supercell yielded a ZPL energy of 779 meV. We regard this value as the final result within the standard HSE06 methodology and assign an uncertainty of order 0.1 eV, consistent with previously reported models of similar systems containing an active carbon dangling bond in silicon~\cite{Udvarhelyi_2022, Dhaliah_2022}. Furthermore, because bound-exciton states can be sensitive to the supercell volume and localization, we also provide an alternative estimate using a finite-size correction and modified exchange functional following Ref.~\onlinecite{Nangoi_2026}. A rough extrapolation based on the 216- and 512-atom supercells yields a dilute-limit estimate of 836 meV for the ZPL energy. A similar systematic shift in the ZPL energies was reported in T-center-like electronic structures using the HSE06 functional~\cite{Nangoi_2026}, with the experimental ZPL energy expected to lie between the values calculated in the 512- and 216-atom models. Our calculations align with this trend. Furthermore, a calculation method using the PBE0 functional with a mixing parameter of 0.136 is proposed to more accurately capture the ZPL energy of T-center-like defects~\cite{Nangoi_2026}. This method results in a 599 meV ZPL energy for the $\text{C}_\text{i}\text{N}_\text{i}$ defect in the 216-atom model. Applying this energy shift to the dilute-limit approximation results in a ZPL energy of 734 meV. This is in line with the HSE06 trends explained above, and matches the experimental N1 ZPL energy.

\begin{figure}[t]
    \centering
    \includegraphics[width=0.95\linewidth]{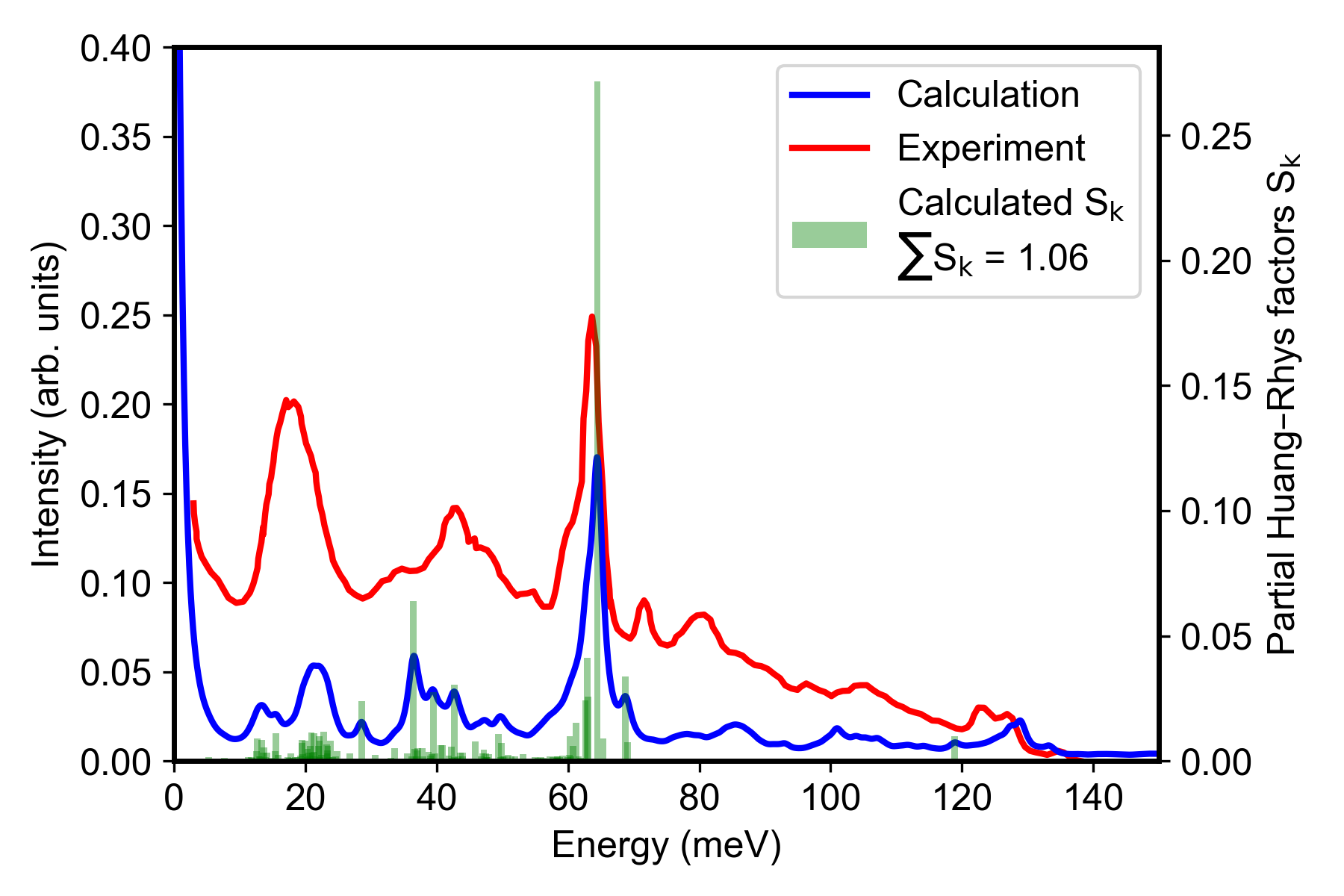}
    \caption{Comparison of the calculated phonon sideband (blue) of the $\text{C}_\text{i}\text{N}_\text{i}$ defect to the experimental sideband of the N1 emitter reported in Ref.~\onlinecite{Dornen_1985} (red). The calculated total Huang-Rhys factor of 1.06 corresponds to a Debye-Waller factor of 0.35.}
    \label{fig:sideband}
\end{figure}

To identify the N1 center, further experimental fingerprints from its optical sideband spectrum are considered. The comparison of the calculated and measured sidebands of the $\text{C}_\text{i}\text{N}_\text{i}$ defect and the N1 center plotted in Fig.~\ref{fig:sideband} shows good agreement. We highlight the exceptionally strong phonon coupling near the $\text{O}^{\Gamma}$ band as a strikingly matching feature. The identified optically active LVMs are at 64.5 meV, 68.8 meV, and 118.9 meV, in close agreement with experiments. The calculated IPR values of 11.0, 11.6, and 1.9, respectively, reveal that the modes around $\text{O}^{\Gamma}$ are only quasi-localized, while the largest energy LVM is an in-plane breathing mode strongly localized on the two interstitial atoms. Based on these findings, we identify the most stable $\text{C}_\text{i}\text{N}_\text{i}$ structure of the carbon-nitrogen pair in silicon as the origin of the N1 ZPL line. Moreover, we calculate a relatively large Debye-Waller factor (DWF) of 0.35 and an optical lifetime of 168 ns for this defect. We note that the calculated lifetime is sensitive to the applied finite-size correction factor. Consequently, the reported value should be regarded as an order-of-magnitude estimate rather than a precise prediction. Within these limitations, the calculated lifetime still compares reasonably with the 940 ns optical lifetime reported for the T center~\cite{Bergeron_2020}. These optical properties make the N1 center a strong contender for quantum photonics among bound-exciton telecom emitters in silicon.

\begin{figure*}[t]
    \centering
    \includegraphics[width=0.95\linewidth]{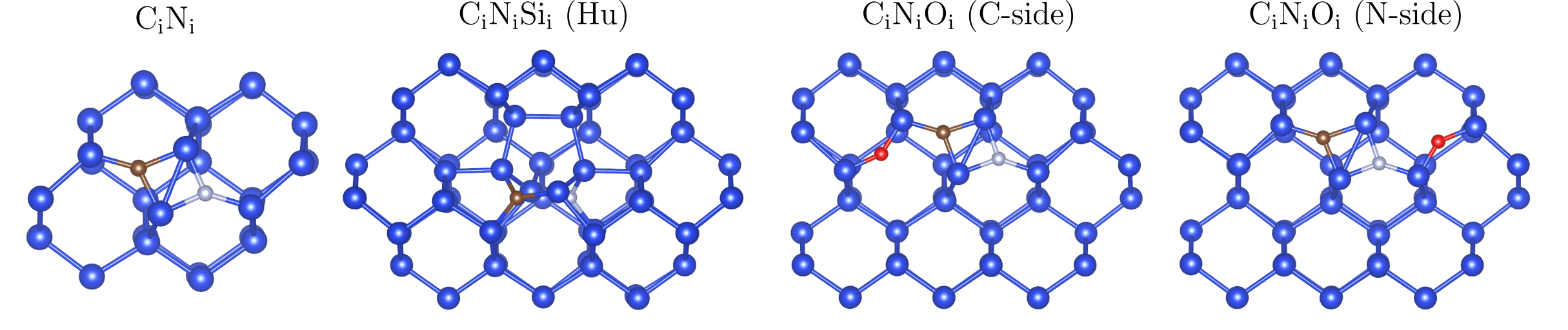}
    \caption{Geometric structures of the calculated interstitial aggregate complexes with the strongest binding energies in their respective atomic constitution. Based on the close agreement between the calculated optical and phonon properties and previous experiments, the $\text{C}_\text{i}\text{N}_\text{i}$ defect is identified as the microscopic origin of the N1 color center. The remaining structures are proposed as viable candidates for further color centers in the N-line series. Dark blue, light blue, brown, and red balls represent Si, N, C, and O atoms, respectively.}
    \label{fig:structures}
\end{figure*}

The calculated ZPL transition energies for the other candidates lie close to that of the $\text{C}_\text{i}\text{N}_\text{i}$ defect. The calculated relative energy shifts ($\Delta \text{ZPL}$) with respect to the $\text{C}_\text{i}\text{N}_\text{i}$ ZPL are 20, 23, and 25 meV for the $\text{C}_\text{i}\text{N}_\text{i}\text{Si}_\text{i}$ (Hu), $\text{C}_\text{i}\text{N}_\text{i}\text{O}_\text{i}$ (C-side), and (N-side), respectively. This shows possible matches with the N2, N4, and N5 line energy order. The assignment of the lines is further supported by the matching point symmetries obtained for the N2 and N5 centers. For the origin of the missing N3 line, a straightforward choice could be the $\text{C}_\text{i}\text{N}_\text{i}\text{Si}_\text{i}\text{O}_\text{i}$ (Hu) and $\text{C}_\text{i}\text{N}_\text{i}\text{Si}_\text{i}\text{O}_\text{i}$ (bc) stuctures, both showing matching symmetry with N3. However, their calculated ZPL energies deviate more substantially from the previously assigned ones ($\Delta \text{ZPL}$ of 0 and 3 meV, respectively). This discrepancy makes it difficult to draw reliable conclusions for the N3 line. The typical absolute uncertainty of hybrid‑functional ZPL calculations is on the order of 0.1 eV, which is comparable to the spacing of the N‑series. While this limits definitive assignments, relative ZPL shifts are expected to be more reliable due to partial cancellation of systematic errors.

Our calculations on the optical properties of the candidate structures conclude with strong evidence for the $\text{C}_\text{i}\text{N}_\text{i}$ defect as the origin of the N1 center. Moreover, all the candidate structures lie in a 25 meV range of the former ZPL energy, suggesting that they are possible origins of the further lines in the N-series. Considering point symmetries and relative stabilities, we propose candidate structures for the N2, N4, and N5 lines. The final candidate models are shown in Fig.~\ref{fig:structures}.

\section{Conclusion}

We applied first-principles calculations on complex defects consisting of carbon, nitrogen, oxygen, and self-interstitial defects in silicon. We find that the neighboring nitrogen-carbon interstitial pair ($\text{C}_\text{i}\text{N}_\text{i}$) combines the smallest formation energy with the largest binding energy among the various structures of carbon-nitrogen pairs in silicon. Its exceptional stability, $\text{C}_{1\text{h}}$ symmetry, electronic structure, ZPL energy, phonon sideband characteristics, and LVM energies all show reasonable agreement with the experimental properties of the N1 color center. Based on multiple matching properties, the identification of the N1 center as the $\text{C}_\text{i}\text{N}_\text{i}$ defect complex is robust. For the remaining N-series lines, we propose a set of feasible candidate structures consistent with the available experimental constraints. Our motif‑based search for an aggregated self‑interstitial near the N1 core yields the optically active $\text{C}_\text{i}\text{N}_\text{i}\text{Si}_\text{i}$ (Hu) configuration, whose symmetry and ZPL shift make it a candidate for the N2 center. Mapping oxygen binding around the N1 core identifies the most stable $\text{C}_\text{i}\text{N}_\text{i}\text{O}_\text{i}$ (N-side) and a higher-energy metastable C-side variant as candidate structures for the N5 and N4 centers, respectively. Their ordering is consistent with the experimental ZPL sequence and with the weaker PL intensity of N4, which suggests a higher formation energy. Calculations on the origin of the N3 line did not result in an unambiguous conclusion. However, we propose the oxygen-modified $\text{C}_\text{i}\text{N}_\text{i}\text{Si}_\text{i}\text{O}_\text{i}$ (Hu) and $\text{C}_\text{i}\text{N}_\text{i}\text{Si}_\text{i}\text{O}_\text{i}$ (bc) stuctures as key candidates for further investigations. Our findings highlight the family of N color centers in silicon as telecom emitters with desirable optical properties. Furthermore, the N centers are isoelectronic to the T center and possess a comparable electronic structure, making them promising alternative qubit candidates. Our calculations provide valuable insight into the atomic structure origin of these color centers, paving the way for their further theoretical and experimental investigation.

\section*{Data availability}
The data supporting the findings of this study are included in the manuscript and its Supplemental Material. Raw input and output files of the calculations are available in the public repository at:
\url{https://doi.org/10.48505/nims.6358}

\section*{Acknowledgement}
P.U. acknowledges the support from the ICYS fellowship and research fund at the National Institute for Materials Science. The helpful discussions with T. Teraji and Y. Yamaji are gratefully acknowledged. This work was achieved through the use of the Supercomputer System SQUID at the D3 Center at the University of Osaka and the Numerical Materials Simulator at the National Institute for Materials Science.

\foreach \x in {1,...,7}
{%
\clearpage
\includepdf[pages={\x}]{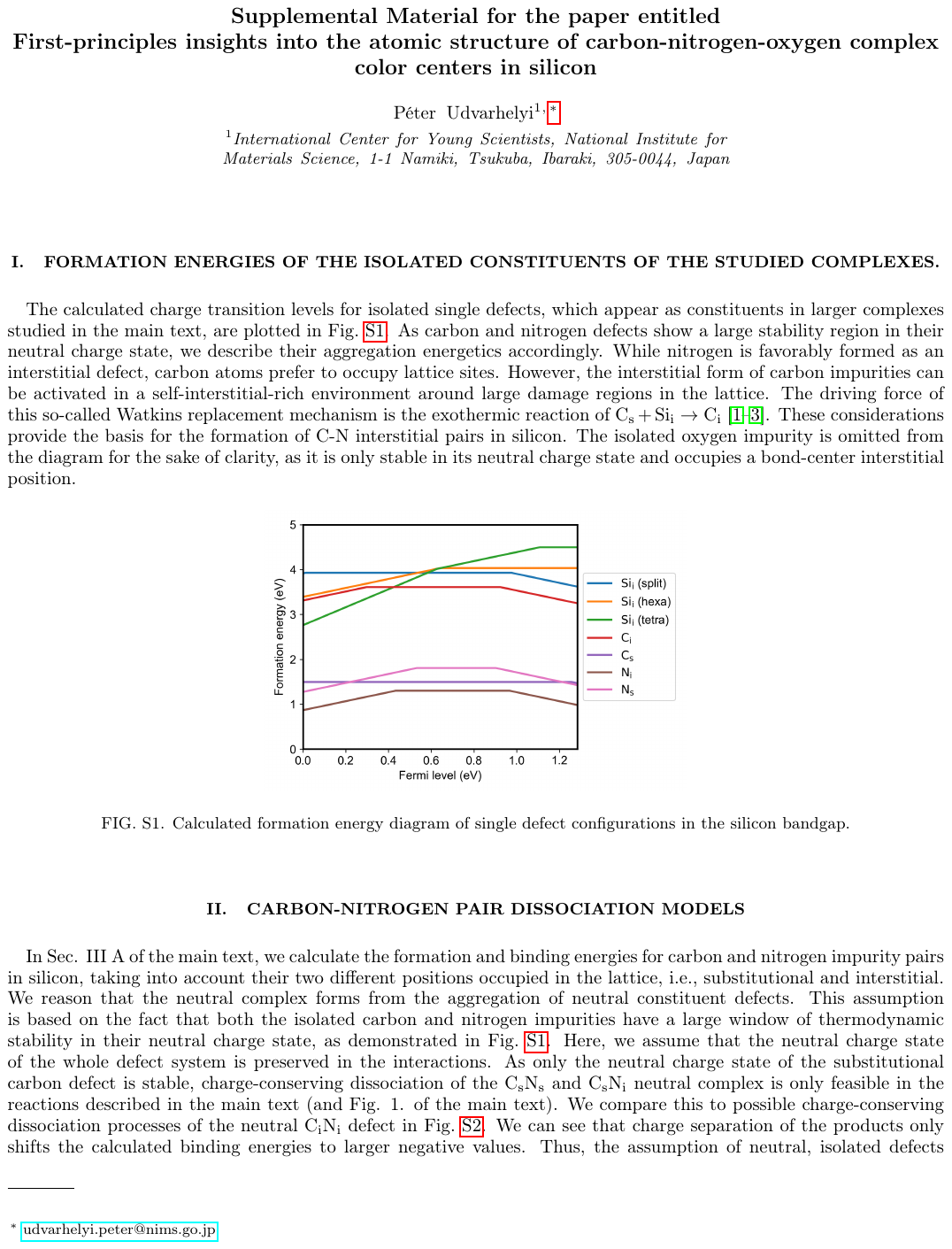}
}
\end{document}